\documentclass[10pt,twocolumn,letterpaper]{article}

\usepackage[pagenumbers]{cvpr} 
\usepackage{graphicx}
\usepackage{amsmath}
\usepackage{amssymb}
\usepackage{booktabs} 
\usepackage[normalem]{ulem}
\useunder{\uline}{\ul}{} 
\usepackage[table]{xcolor}
\usepackage{colortbl} 
\usepackage{multirow} 
\usepackage{booktabs}

\usepackage{etoolbox}
\usepackage{silence}
\makeatletter
\robustify\@latex@warning@no@line
\makeatother
\usepackage{authblk}

\renewcommand*{\Affilfont}{\normalsize}

\makeatletter
\renewcommand\maketitle{\AB@maketitle} 
\renewcommand\AB@affilsepx{\quad\protect\Affilfont} 
\makeatother 

\usepackage{fancyhdr}

\usepackage[pagebackref,breaklinks,colorlinks]{hyperref}

\usepackage[capitalize]{cleveref}
\crefname{section}{Sec.}{Secs.}
\Crefname{section}{Section}{Sections}
\Crefname{table}{Table}{Tables}
\crefname{table}{Tab.}{Tabs.}

\def\cvprPaperID{*****} 
\def\confName{CVPR}
\def\confYear{2022}

\title{Restore-RWKV: Efficient and Effective Medical Image Restoration with RWKV}

\author[1]{Zhiwen Yang} 
\author[1]{Jiayin Li}
\author[2]{Hui Zhang}
\author[3]{Dan Zhao} 
\author[4]{Bingzheng Wei}
\author[1]{Yan Xu\thanks{ Corresponding author (xuyan04@gmail.com)}}
\affil[1]{Beihang University}
\affil[2]{Tsinghua University}
\affil[3]{Peking Union Medical College} 
\affil[4]{ByteDance Inc.}

\begin{document}
\maketitle

\begin{abstract}
Transformers have revolutionized medical image restoration, but the quadratic complexity still poses limitations for their application to high-resolution medical images. The recent advent of the Receptance Weighted Key Value (RWKV) model in the natural language processing field has attracted much attention due to its ability to process long sequences efficiently. To leverage its advanced design, we propose Restore-RWKV, the first RWKV-based model for medical image restoration. Since the original RWKV model is designed for 1D sequences, we make two necessary modifications for modeling spatial relations in 2D medical images. First, we present a recurrent WKV (Re-WKV) attention mechanism that captures global dependencies with linear computational complexity. Re-WKV incorporates bidirectional attention as basic for a global receptive field and recurrent attention to effectively model 2D dependencies from various scan directions. Second, we develop an omnidirectional token shift (Omni-Shift) layer that enhances local dependencies by shifting tokens from all directions and across a wide context range. These adaptations make the proposed Restore-RWKV an efficient and effective model for medical image restoration. Even a lightweight variant of Restore-RWKV, with only 1.16 million parameters, achieves comparable or even superior results compared to existing state-of-the-art (SOTA) methods. Extensive experiments demonstrate that the resulting Restore-RWKV achieves SOTA performance across a range of medical image restoration tasks, including PET image synthesis, CT image denoising, MRI image super-resolution, and all-in-one medical image restoration. Code is available at: \href{https://github.com/Yaziwel/Restore-RWKV.git}{https://github.com/Yaziwel/Restore-RWKV}.
\end{abstract}

\section{Introduction}
\label{sec_intro}
Medical image restoration (MedIR) aims at recovering the high-quality (HQ) medical image from its degraded low-quality (LQ) counterpart. It encompasses a variety of tasks such as PET image synthesis \cite{xiang2017xiang,xu2017200x,chan2018dcnn,zhou2020cyclewgan,zhou2022sgsgan,luo2022argan,jang2023spachtransformer,yang2023drmc}, CT image denoising \cite{chen2017redcnn,yang2018wgan_vgg,liang2020edcnn,zhang2021transct,luthra2021eformer,wang2023ctformer,ozturk2024denomamba}, MRI image super-resolution \cite{yang2017dagan,chen2020fawdn,huang2022swinmr,huang2022sdaut,sun2025funet}, and all-in-one medical image restoration \cite{li2022airnet,yang2023drmc,yang2024amir,kong2024mio}. This field is particularly challenging due to the ill-posed nature of the problem, where crucial information about the image content is often lost in LQ images. Recent years have witnessed significant progress in MedIR, largely driven by advanced deep learning models such as convolutional neural networks (CNNs) \cite{lecun1989cnn,he2016resnet}, Transformers \cite{vaswani2017transformer,dosovitskiy2020vit}, and Mambas \cite{gu2023mamba,zhu2024vision-mamba,ozturk2024denomamba,huang2025mambamir}.

\begin{figure}[t]
\centering
\includegraphics[width=0.48\textwidth]{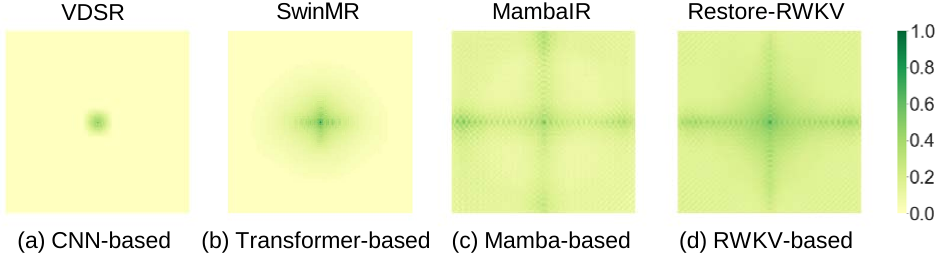}
\caption{The effective receptive field (ERF) \cite{luo2016erf} visualization for different efficient models. A more extensively distributed dark area indicates a larger ERF. Our proposed Restore-RWKV achieves the most significant global ERF.}
\label{fig_erf}
\end{figure} 

Despite significant advancements, current CNN-based, Transformer-based, and Mamba-based models face challenges in striking a proper balance between a large receptive field and computational efficiency. To some extent, improvements in medical image restoration performance are largely dependent on the model's receptive field. On one hand, a larger receptive field enables the model to capture information from a wider region, allowing it to reference more pixels to improve the recovery of the degraded areas. On the other hand, a larger receptive field allows the model to capture high-level patterns and structures, which are essential for effective medical image restoration \cite{guo2025mambair}. However, increasing the receptive field often leads to higher computational complexity in model design. Among existing studies, CNN-based models \cite{kim2016vdsr} often face constraints in their effective receptive field \cite{luo2016erf} (ERF, as illustrated in Fig.~\ref{fig_erf} (a)) due to the limited kernel size of the convolution operator. This limitation hampers their ability to capture information across broader ranges to compensate for lost information in degraded pixels. In contrast, Transformer-based models excel in modeling long-range dependencies and achieving a global ERF through self-attention \cite{dosovitskiy2020vit}. However, the computational complexity of self-attention grows quadratically with spatial resolution, rendering standard Transformers impractical for high-resolution medical images. To mitigate this, many other Transformer-based models aim to reduce computational demands by restricting attention to smaller windows \cite{huang2022swinmr}, inevitably diminishing the ERF (as depicted in Fig.~\ref{fig_erf} (b)) and conflicting with the goal of capturing long-range dependencies. Mamba-based models \cite{guo2025mambair,ji2024mamba_mri,ozturk2024denomamba} serve as an efficient alternative to Transformers, leveraging state space models (SSMs) \cite{gu2023mamba} to causally capture long-range dependencies with linear computational complexity. Recent studies suggest that Mamba-based models can achieve performance comparable to or even better than Transformer-based models. However, their inherent unidirectional sequence modeling in SSMs still poses challenges in achieving an optimal ERF for 2D images (as shown in Fig.~\ref{fig_erf} (c)). In summary, existing models in MedIR struggle to effectively address the trade-off between a global receptive field and computational efficiency.

The \textbf{R}eceptance \textbf{W}eighted \textbf{K}ey \textbf{V}alue (RWKV) \cite{peng2023rwkv,peng2024rwkv6} model, originating from natural language processing (NLP), is emerging as an alternative to the Transformer. RWKV \cite{peng2023rwkv} introduces two significant innovations. First, it introduces a WKV attention mechanism that achieves long-range dependencies with linear computational complexity, addressing the expensive computational cost of self-attention in Transformers. Second, it incorporates a token shift layer to enhance the capture of local dependencies, which is typically overlooked by the standard Transformer \cite{vaswani2017transformer,dosovitskiy2020vit}. With these advanced designs, RWKV demonstrates strong capacity and scalability, performing admirably with large-scale NLP \cite{peng2023rwkv,peng2024rwkv6} and vision datasets \cite{duan2024vrwkv,fei2024rwkv-diffusion,he2024pointrwkv,gu2024rwkv-clip,yuan2024rwkv-sam}. Recent studies \cite{peng2024rwkv6,he2024pointrwkv,yuan2024rwkv-sam} have shown that RWKV-based models outperform Transformer-based models and linearly scaled Mamba-based models in both effectiveness and efficiency. Despite pioneering efforts to adopt RWKV for vision tasks, it is still in the initial stages of extending the original RWKV from modeling 1D sequences to modeling 2D images, and the adaptation of RWKV to the field of MedIR still remains unexplored.


In this paper, we propose Restore-RWKV, an efficient and effective model adapted from RWKV for medical image restoration. The original RWKV \cite{peng2023rwkv} is designed for handling 1D sequences and has limitations in capturing spatial dependencies in 2D medical images. To address these limitations, we make essential modifications to two key components of RWKV—the attention layer and the token shift layer—to adapt it for processing 2D medical images. (1) We introduce a recurrent WKV (Re-WKV) attention mechanism that effectively captures global dependencies with linear computational complexity. Unlike the causal WKV attention in the original RWKV, which is unidirectional and has a limited receptive field, Re-WKV utilizes a bidirectional attention mechanism to achieve a global receptive field. Additionally, it employs a recurrent attention mechanism that effectively models 2D dependencies across various scan directions. (2) We develop an omnidirectional token shift (Omni-Shift) layer to accurately capture local dependencies. Previous token shift layers only shift adjacent tokens from limited directions using simple interpolation.  In contrast, Omni-Shift enhances interaction among neighbors by shifting tokens from all directions and over a large context range through efficient depth-wise convolution. Furthermore, during training, Omni-Shift employs a structural re-parameterization strategy to learn to accurately shift tokens from various context ranges, while maintaining the original structure for efficiency during testing. With these two innovations, the proposed Restore-RWKV effectively models both global and local dependencies in 2D images, achieving the largest global receptive field (as shown in Fig.~\ref{fig_erf} (d)). Extensive experimental results indicate that our proposed Restore-RWKV can achieve superior performance with good efficiency, serving as a general restoration backbone for various tasks, including MRI image super-resolution, CT image denoising, PET image synthesis, and all-in-one medical image restoration.

Our main contribution can be summarized  as follows:
\begin{itemize}
\item  We propose Restore-RWKV, which pioneers the adaptation of the RWKV model for medical image restoration. It has proven an efficient and effective alternative for medical image restoration backbones.

\item  We present a recurrent WKV (Re-WKV) attention mechanism that effectively captures global dependencies in high-resolution medical images with linear computational complexity.

\item  We develop an omnidirectional token shift (Omni-Shift) layer, enhancing local dependencies by establishing accurate token interactions from all directions.
\end{itemize}

\section{Related Work} 
\subsection{Medical Image Restoration} 
High-quality (HQ) medical images, which are essential for accurate disease diagnosis, typically offer superior image quality but come with the drawbacks of longer scanning times or increased radiation exposure. In contrast, low-quality (LQ) images, which reduce scanning time and radiation exposure, often suffer from degraded image quality. As a result, there is significant interest in medical image restoration (MedIR) techniques aimed at recovering HQ medical images from their degraded LQ counterparts. Typical MedIR tasks include PET image synthesis \cite{xiang2017xiang,xu2017200x,chan2018dcnn,zhou2020cyclewgan,zhou2022sgsgan,luo2022argan,jang2023spachtransformer,yang2023drmc}, CT image denoising \cite{chen2017redcnn,yang2018wgan_vgg,liang2020edcnn,zhang2021transct,luthra2021eformer,wang2023ctformer,ozturk2024denomamba}, MRI image super-resolution \cite{yang2017dagan,chen2020fawdn,huang2022swinmr,huang2022sdaut,sun2025funet}, and all-in-one medical image restoration for multi-task MedIR\cite{li2022airnet,yang2023drmc,yang2024amir,kong2024mio}. Recently, deep learning-based models have become the primary approach for addressing MedIR tasks. These methods can be broadly categorized into CNN-based models, Transformer-based models, and Mamba-based models.

\textbf{CNN-based Models.} Over the past decades, CNN-based models \cite{chen2017redcnn,xiang2017xiang,luo2022argan} have made extraordinary contributions to the field of MedIR due to their strong representation capabilities. A key advantage of CNN-based models for MedIR is their ability to capture local context and build local dependencies through convolution. This allows the neighborhood of a degraded pixel to be used as a reference to aid in its recovery. However, due to the limited kernel size of convolutions, CNN-based methods often suffer from restricted receptive fields and cannot model long-range dependencies. While some approaches have tried to expand the receptive field by designing deeper networks \cite{kim2016vdsr,lim2017edsr} or utilizing the multi-scale U-Net \cite{chen2017redcnn,luo2022argan} architecture, these improvements are still limited and struggle to capture long-range dependencies effectively. 

\begin{figure*}[t]
\centering
\includegraphics[width=0.9\textwidth]{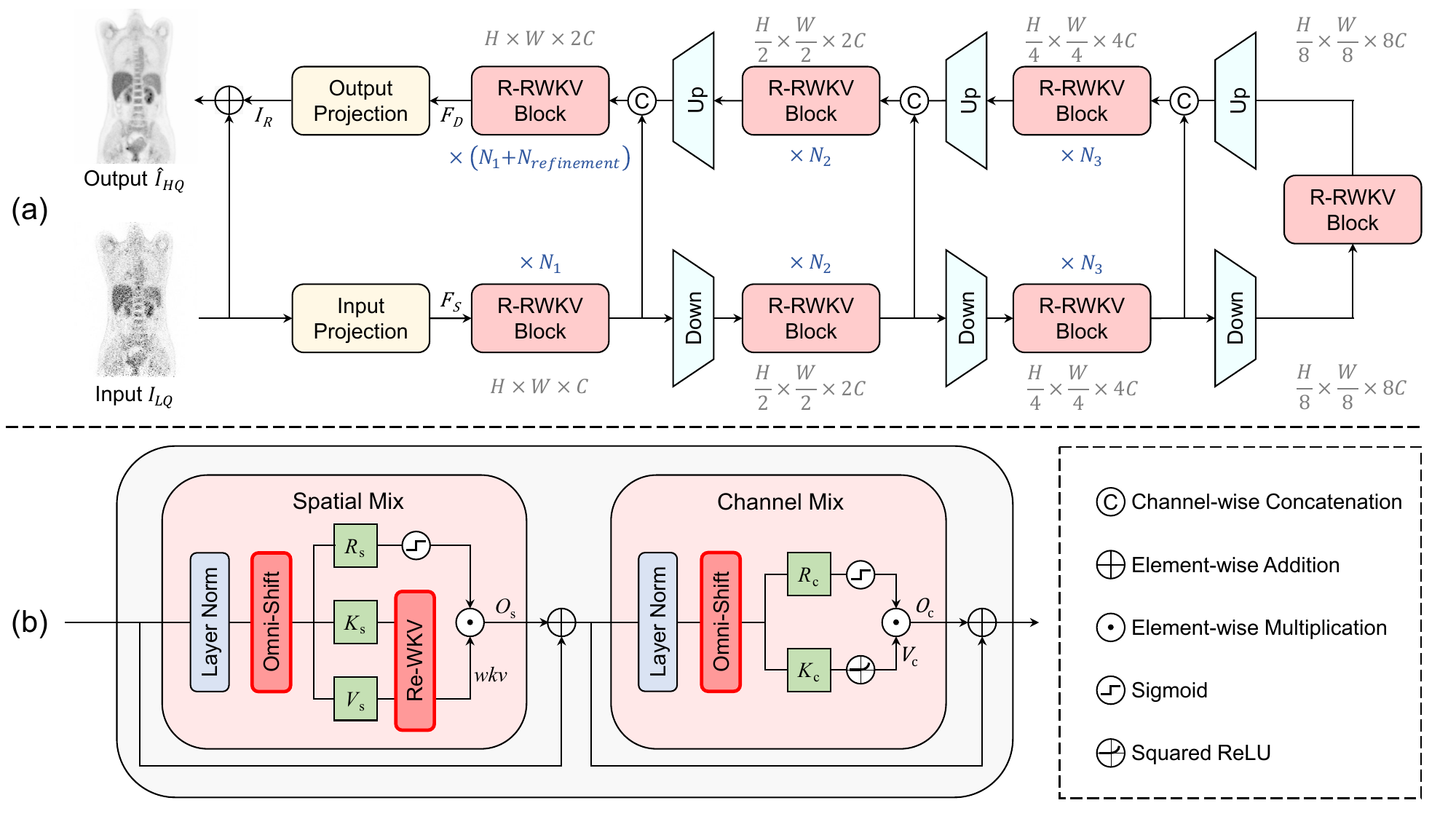}
\caption{(a) Overview of the Restore-RWKV architecture. (b) Illustration of the R-RWKV block, which incorporates a Re-WKV attention mechanism to model global dependencies with linear complexity, and an Omni-Shift layer to capture local context.}
\label{fig_framework}
\end{figure*} 

\textbf{Transformer-based Models.} The Transformer model \cite{vaswani2017transformer,dosovitskiy2020vit}, originally developed for natural language processing (NLP), has been adapted to achieve significant performance in vision tasks, including MedIR. Compared to CNN-based models, Transformer-based models excel at building long-range dependencies through global self-attention. This capability allows Transformer-based models to better utilize information from the entire image to compensate for the information lost in degraded pixels. However, the standard Transformer architecture suffers from quadratic computational complexity due to its self-attention mechanism, limiting its practicality for MedIR tasks involving high-resolution images. To address this issue, several approaches \cite{liang2021swinir,chen2023sparsevit,tu2024ipt-v2} have explored efficient attention mechanisms by limiting the scope of token interactions in attention calculation, such as window attention \cite{liang2021swinir} and sparse attention \cite{chen2023sparsevit}. Nevertheless, these efficient attention mechanisms often come at the cost of a reduced global receptive field and struggle to balance the trade-off between accuracy and efficiency.

\textbf{Mamba-based Models.} Mamba \cite{gu2023mamba,zhu2024vision-mamba}, a sequence model backbone grounded in state-space models (SSMs), has emerged as a prominent alternative to Transformers. Unlike Transformers, which exhibit quadratic computational complexity, Mamba-based models can model long-range dependencies with linear computational complexity. This capability allows Mamba-based models to potentially capture global dependencies in high-resolution medical images through SSMs without the need for efficient designs like window partitioning. Recent studies have shown that Mamba-based models \cite{guo2025mambair,ozturk2024denomamba,huang2025mambamir} outperform Transformer-based models in MedIR tasks. However, due to the unidirectional causal attention mechanism of SSM, its receptive field is limited to the range from the first token to the current token in the sequence. As a result, it fails to capture information from subsequent tokens and cannot achieve a global receptive field. While employing a multi-directional scanning strategy \cite{liu2024vmamba,guo2025mambair} can expand the receptive field to encompass a global context, it proves neither efficient nor effective. Consequently, despite Mamba's ability to model long-range dependencies, it still faces limitations in capturing global dependencies in 2D medical images.

\subsection{Receptance Weighted Key Value}
The \textbf{R}eceptance \textbf{W}eighted \textbf{K}ey \textbf{V}alue (RWKV) model \cite{peng2023rwkv,peng2024rwkv6}, which originates from the field of NLP, is an emerging efficient alternative to Transformers. Compared to the standard Transformer \cite{vaswani2017transformer}, RWKV offers two significant advantages. First, RWKV proposes a WKV attention mechanism to build long-range dependencies with linear computational complexity, addressing the quadratic computational complexity of self-attention in Transformers. Second, RWKV introduces a token shift layer to capture the local context, which is often ignored by the standard Transformer. Recent study \cite{peng2024rwkv6} indicates that RWKV can achieve performance comparable to or even better than that of both Transformers and Mamba in NLP tasks. Recently, Vision-RWKV \cite{duan2024vrwkv} has transferred RWKV from NLP to vision tasks, demonstrating superior performance compared to vision Transformers with reduced computational complexity. To better utilize spatial information in 2D images, Vision-RWKV proposes a bidirectional WKV attention mechanism to capture global dependencies and a quad-directional token shift mechanism to capture local context from four different directions. Building on RWKV and Vision-RWKV, several RWKV-based models have been developed for various vision-related tasks, such as Diffusion-RWKV \cite{fei2024rwkv-diffusion} for image generation, RWKV-SAM \cite{yuan2024rwkv-sam} for segment anything, Point-RWKV \cite{he2024pointrwkv} for 3D point cloud learning, and RWKV-CLIP \cite{gu2024rwkv-clip} for vision-language representation learning. However, few studies have validated the performance of RWKV in low-level vision tasks like medical image restoration. In this paper, we reveal that current RWKV-based models still have limitations in building both global and local dependencies in 2D images, hindering their performance in MedIR tasks. To address this, we propose Restore-RWKV, which efficiently and effectively models both global and local dependencies, achieving superior performance in restoring medical images.

\section{Method} 
We aim to build an efficient and effective model for medical image restoration based on the \textbf{R}eceptance \textbf{W}eighted \textbf{K}ey \textbf{V}alue (RWKV) \cite{peng2023rwkv}, termed Restore-RWKV. In Section~\ref{sec_network_arch}, we introduce the network architecture of Restore-RWKV, and in Section~\ref{sec_r-rwkv-block}, we present a novel R-RWKV block as its fundamental component for feature extraction. Since the original RWKV \cite{peng2023rwkv} is designed for processing 1D sequences, we incorporate two innovations in the R-RWKV block to adapt it for capturing global and local dependencies in 2D images. First, in Section~\ref{sec_re-wkv}, we introduce a recurrent WKV (Re-WKV) attention mechanism that effectively builds global dependencies in 2D images with linear computational complexity. Second, in Section~\ref{sec_omni-shift}, we introduce an omnidirectional token shift mechanism that accurately captures local context to enhance the model performance. We will elaborate on the details of Restore-RWKV in the following section.

\subsection{Restore-RWKV Architecture} 
\label{sec_network_arch}
As shown in Fig~\ref{fig_framework} (a), the proposed Restore-RWKV is a 4-level U-shaped encoder-decoder architecture with 3 times downsampling and upsampling, which enjoys the advantage of capturing image features at different hierarchies and computational efficiency \cite{ronneberger2015unet}. Given an input low-quality (LQ) image $I_{LQ} \in \mathbb{R}^{H\times W \times 1}$, Restore-RWKV first employs a $3 \times 3$ convolutional layer as an input projection to project the input image to a shallow feature $F_{S} \in \mathbb{R}^{H\times W \times C}$, where $H$, $W$, and $C$ denote the height, width, and channel, respectively. Then the shallow feature $F_{S}$ undergoes a 4-level hierarchical encoder-decoder and is transformed into a deep feature $F_{D} \in \mathbb{R}^{H\times W \times 2C}$. Each level of the encoder-decoder comprises $N_i$, where $i \in \{1,2,3,4\}$, R-RWKV blocks for feature extraction, with an additional $N_{\text{refinement}}$ R-RWKV blocks used for feature refinement in the final decoder level. For feature downsampling, Restore-RWKV utilizes a $1\times1$ convolutional layer and a pixel-unshuffle operation \cite{shi2016pixelshuffle}, reducing the spatial size by half and doubling the channel number. For feature upsampling, Restore-RWKV employs a pixel-shuffle operation \cite{shi2016pixelshuffle} and a $1\times1$ convolutional layer, doubling the spatial size and halving the channel number. To aid in the restoration process, encoder features are concatenated with decoder features via skip connections. The deep feature $F_{D}$ is then projected out to a residual image $I_R \in \mathbb{R}^{H\times W \times 1}$ by a $3\times3$ convolutional layer. Finally, the restored image $\hat{I}_{HQ}$ can be obtained by adding the input $I_{LQ}$ and the residual image $I_{R}$:  $\hat{I}_{HQ} = I_{LQ} + I_{R}$.

\subsection{R-RWKV Block} 
\label{sec_r-rwkv-block}

The R-RWKV blocks play a crucial role in feature extraction across different hierarchical levels within Restore-RWKV. As shown in Fig.~\ref{fig_framework} (b), our proposed R-RWKV block follows the design of the original RWKV block \cite{peng2023rwkv} and integrates a spatial mix module and a channel mix module, enabling spatial-wise token interaction and channel-wise feature fusion, respectively.  Since the original RWKV block \cite{peng2023rwkv} is tailored for processing 1D sequences and lacks comprehensive capability in capturing global and local contexts in 2D images, our R-RWKV block introduces two innovations to tackle the increased dimensionality of 2D images: recurrent WKV attention (Re-WKV) for capturing global dependencies and omnidirectional token shift (Omni-Shift) for capturing local context. The data flow in R-RWKV is detailed as follows.

\textbf{Spatial Mix.} The spatial mix module is designed to build long-range dependencies within tokens across the spatial dimension. Given an input feature flattened to a one-dimensional sequence $X \in \mathbb{R}^{T \times C}$, where $T=H \times W$ denotes the total number of tokens, the spatial mix module first passes it through a layer normalization (LN) and an Omni-Shift layer (refer to Section~\ref{sec_omni-shift}):
\begin{equation}
X_s = \operatorname{Omni-Shift}(\operatorname{LN}(X)).
\end{equation}
Here, LN is applied to stabilize the training process. Our proposed Omni-Shift is specifically introduced to capture local context and expand the context range of individual tokens. Then $X_s$ is passed through three parallel linear projection layers to obtain the matrices of receptance $R_s \in \mathbb{R}^{T \times C}$, key $K_s \in \mathbb{R}^{T \times C}$, and value $V_s \in \mathbb{R}^{T \times C}$:
\begin{equation}
R_s=X_s W_{R_s}, \quad K_s=X_s W_{K_s}, \quad V_s=X_s W_{V_s},
\end{equation}  
where $W_{R_s}$, $W_{K_s}$, and $W_{V_s}$ represent the three linear projection layers. Subsequently, $K_s$ and $V_s$ are utilized  to acquire the global attention result $wkv \in \mathbb{R}^{T \times C}$ by our proposed linear-complexity Re-WKV attention mechanism (as detailed in Section~\ref{sec_omni-shift}): 
\begin{equation}
wkv = \operatorname{Re-WKV}(K_s, V_s).
\end{equation}  

Finally, the receptance after gating $\sigma(R_s)$ modulates the received probability of the attention result $wkv$ through element-wise multiplication:
\begin{equation}
O_s = (\sigma(R_s)\odot wkv)W_{O_s},
\end{equation}  
where $O_s$ denotes the output, $\sigma(\cdot)$ represents the Sigmoid gating function, and $W_{O_s}$ signifies the linear projection layer for output projection.

\textbf{Channel Mix.} The channel mix module aims at performing feature fusion in the channel dimension. Given an input feature $X \in \mathbb{R}^{T \times C}$, the channel mix passes it through LN and Omni-Shift layer as spatial mix:
\begin{equation}
X_c = \operatorname{Omni-Shift}(\operatorname{LN}(X)).
\end{equation} 

Then the receptance $R_c \in \mathbb{R}^{T \times C}$, key $K_c \in \mathbb{R}^{T \times C}$, and value $V_c \in \mathbb{R}^{T \times C}$ can be acquired as follows:
\begin{equation}
R_c=X_c W_{R_c}, \quad K_c=X_c W_{K_c}, \quad V_c=\gamma (K_c) W_{V_c}.
\end{equation}  
Here, $W_{R_c}$, $W_{K_c}$, and $W_{V_c}$ represent the three linear projection layers. $\gamma (\cdot)$ denotes the squared ReLU activation function, known for its enhanced nonlinearity. Notably, $V_c$ is estimated from $K_c$ rather than directly from $X_c$, which differs from the approach used in the spatial mix module. In fact, the transformation from $X_c$ to $K_c$ to $V_c$ involves a multi-layer perception (MLP) consisting of $W_{K_c}$, $\gamma (\cdot)$, and $W_{V_c}$, facilitating channel-wise feature fusion. 

Finally, the output $O_c$ is derived by multiplying $V_c$ with $\sigma(R_c)$ to control the received probability of $V_c$:
\begin{equation}
O_c = (\sigma(R_c)\odot V_c)W_{O_c},
\end{equation}  
where $W_{O_c}$ signifies the linear projection layer for output projection.

\textbf{Differences with RWKV Block.} Our principle is to retain the advantages of the original RWKV block architecture \cite{peng2023rwkv}, which is initially designed for processing 1D sequences, while making essential modifications to accommodate modeling spatial relationships in the context of 2D images. Consequently, the proposed R-RWKV block is built by replacing only the attention layer and token shift layer of the original RWKV block with our proposed Re-WKV attention and Omni-Shift, keeping other layers unchanged. This modification allows R-RWKV to effectively capture global (via Re-WKV) and local (via Omni-Shift) dependencies within the spatial dimensions of 2D images.

\begin{figure*}[t]
\centering
\includegraphics[width=0.85\textwidth]{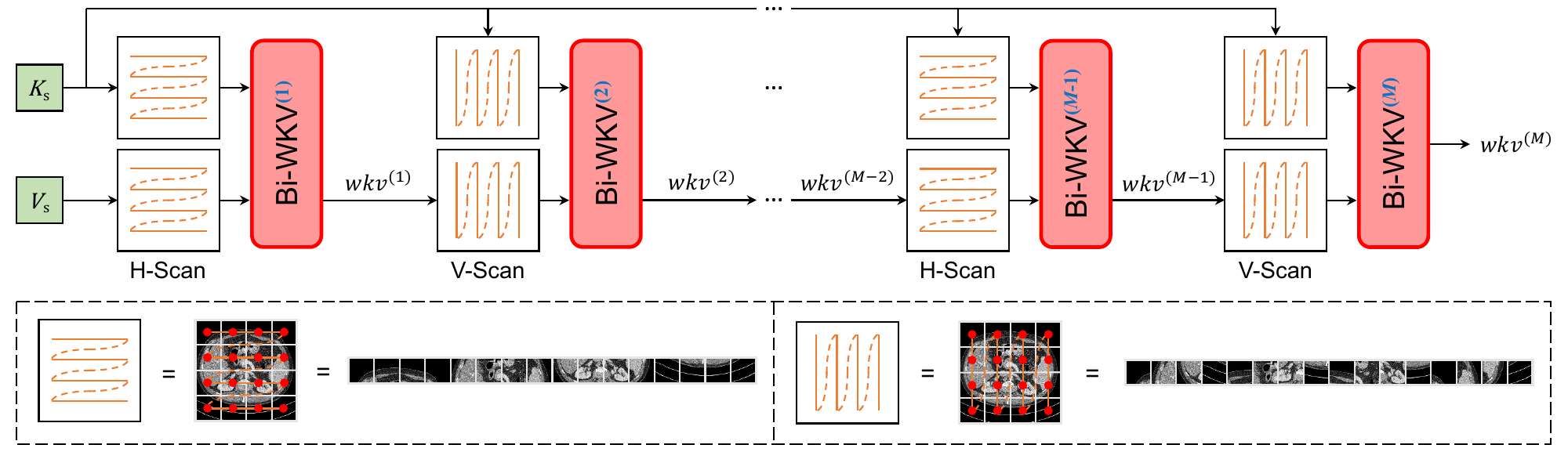}
\caption{Illustrations of the Re-WKV attention mechanism. Re-WKV employs Bi-WKV \cite{duan2024vrwkv} as its basic attention operator and applies Bi-WKV attention to 2D images recurrently through various scan directions to better model global dependencies.}
\label{fig_re_wkv}
\end{figure*} 

\subsection{Recurrent WKV Attention} 
\label{sec_re-wkv} 
The WKV attention mechanism is the core component in the spatial mix to achieve long-range dependencies with linear computational complexity, addressing the quadratic complexity of the standard self-attention in Transformers. However, the original WKV attention in RWKV \cite{peng2023rwkv} is unidirectional (Uni-WKV) and has a limited receptive field confined to the scanned part of sequential data. While this characteristic is well-suited for 1D causal sequence modeling in natural language, it faces challenges when applied to non-causal 2D images that require global modeling. To address these challenges, we propose a recurrent WKV attention (Re-WKV) mechanism for processing 2D images. Re-WKV integrates a basic bidirectional attention mechanism to achieve a global receptive field and a recurrent attention mechanism to better adapt the basic bidirectional attention for modeling 2D image dependencies from multiple scan directions. We introduce the proposed Re-WKV by sequentially detailing its bidirectional and recurrent attention mechanisms.

\textbf{Bidirectional Attention}. To address the limited receptive field of Uni-WKV attention in RWKV \cite{peng2023rwkv}, which spans from the first token to only the current token in a sequence, we follow Vision-RWKV \cite{duan2024vrwkv} by adopting a bidirectional WKV (Bi-WKV) attention mechanism that ensures a global receptive field spanning from the first token to the last token. Given the input projections of key $K_s$ and value $V_s$, the attention result of the current $t$-th token (denoted as $wkv_t  \in \mathbb{R}^{C}$) can be formulated as follows:
\begin{equation}
\begin{aligned}
w k v_t&=\operatorname{Bi-WKV}(K_s, V_s)_t \\
&=\frac{\sum_{i=1, i \neq t}^{T} e^{-(|t-i|-1) / T \cdot w+k_i} v_i+e^{u+k_t} v_t}{\sum_{i=1, i \neq t}^{T} e^{-(|t-i|-1) / T \cdot w+k_i}+e^{u+k_t}}.
\end{aligned}
\label{eq:bi-wkv}
\end{equation}
Here, $T$ denotes the total number of tokens. $k_i \in \mathbb{R}^{C}$  and $v_i \in \mathbb{R}^{C}$ indicate $i$-th spatial token of $K_s$ and $V_s$, respectively. $-(|t-i|-1)/T$ represents the relative position bias between the $t$-th and $i$-th tokens, with a learnable parameter $w \in \mathbb{R}^{C}$ controlling its magnitude. The learnable parameter $u \in \mathbb{R}^{C}$ is a special case of $w$, giving a bonus to the current $t$-th token. To sum up, Eq.~\ref{eq:bi-wkv} indicates that the attention result $wkv_t$ of the $t$-th token is a weighted sum of $V_s$ along the token dimension from $1$ to $T$, with summation weight being collectively determined by relative position bias $-(|t-i|-1)/T$ and key $k_i$.

Bi-WKV ensures both global receptive field and computational efficiency. On the one hand, since the attention result for each token is determined by all other tokens, Bi-WKV guarantees a global receptive field. On the other hand, it eliminates the query-key matrix multiplication, thus avoiding the quadratic computational complexity inherent in standard self-attention mechanisms. According to Vision-RWKV \cite{duan2024vrwkv}, given the input $K_s$ and $V_s$ with the shape of $T \times C$, the practical computational complexity of Bi-WKV is $O(T \times C)$, which scales linearly with the number of tokens $T$.

\textbf{Recurrent Attention}. Bi-WKV attention can still be challenging to apply to 2D images due to its direction-sensitive nature. According to Eq.~\ref{eq:bi-wkv}, Bi-WKV is partially determined by the relative position bias between tokens, indicating that Bi-WKV can be sensitive to the arrangement order of sequential tokens. However, the sequential order of 2D image tokens can vary with different scan directions. Therefore, previous approaches that use a single-direction scan cannot effectively model the dependencies in 2D images. To overcome this limitation, we propose a recurrent WKV attention (Re-WKV) that applies Bi-WKV attention along different scan directions. The mechanism of our proposed Re-WKV is illustrated in Fig.~\ref{fig_re_wkv}, and its core can be formulated as follows:
\begin{equation}
wkv^{(j)} = \operatorname{Bi-WKV}^{(j)}(\operatorname{\Delta}_{dir}(K_s), \operatorname{\Delta}_{dir}(wkv^{(j-1)})).
\label{eq:re-wkv-recurrent}
\end{equation}
Here, $\operatorname{Bi-WKV}^{(j)}(\cdot)$ denotes the $j$-th Bi-WKV attention. $\operatorname{\Delta}_{dir}(\cdot)$ represents the changing direction operation. In two adjacent Bi-WKV attention operations, we alternately use two different scan directions: horizontal scan (H-Scan) and vertical scan (V-Scan). $wkv^{(j)}$ and $wkv^{(j-1)}$ are attention results of the $j$-th and $(j-1)$-th Bi-WKV, respectively. Note that $wkv^{(0)}$ is the input $V_s$. Eq.~\ref{eq:re-wkv-recurrent} indicates the current $j$-th Bi-WKV attention takes the previous $(j-1)$-th Bi-WKV attention result as its input value, complementing the $(j-1)$-th attention result from a different scan direction. The final Re-WKV attention result $wkv$ is obtained after applying Bi-WKV attention recurrently $M$ times:
\begin{equation}
wkv = \operatorname{Re-WKV}(K_s, V_s) = wkv^{(M)}.
\label{eq:re-wkv-final}
\end{equation} 

In comparison with Bi-WKV, Re-WKV enhances global token interactions in 2D images by recurrently performing attention along different scan directions. Furthermore, since $M \ll T$ in our implementation, Re-WKV maintains the linear computational complexity as Bi-WKV.

\subsection{Omnidirectional Token Shift} 
\label{sec_omni-shift}

\begin{figure}[t]
\centering
\includegraphics[width=0.48\textwidth]{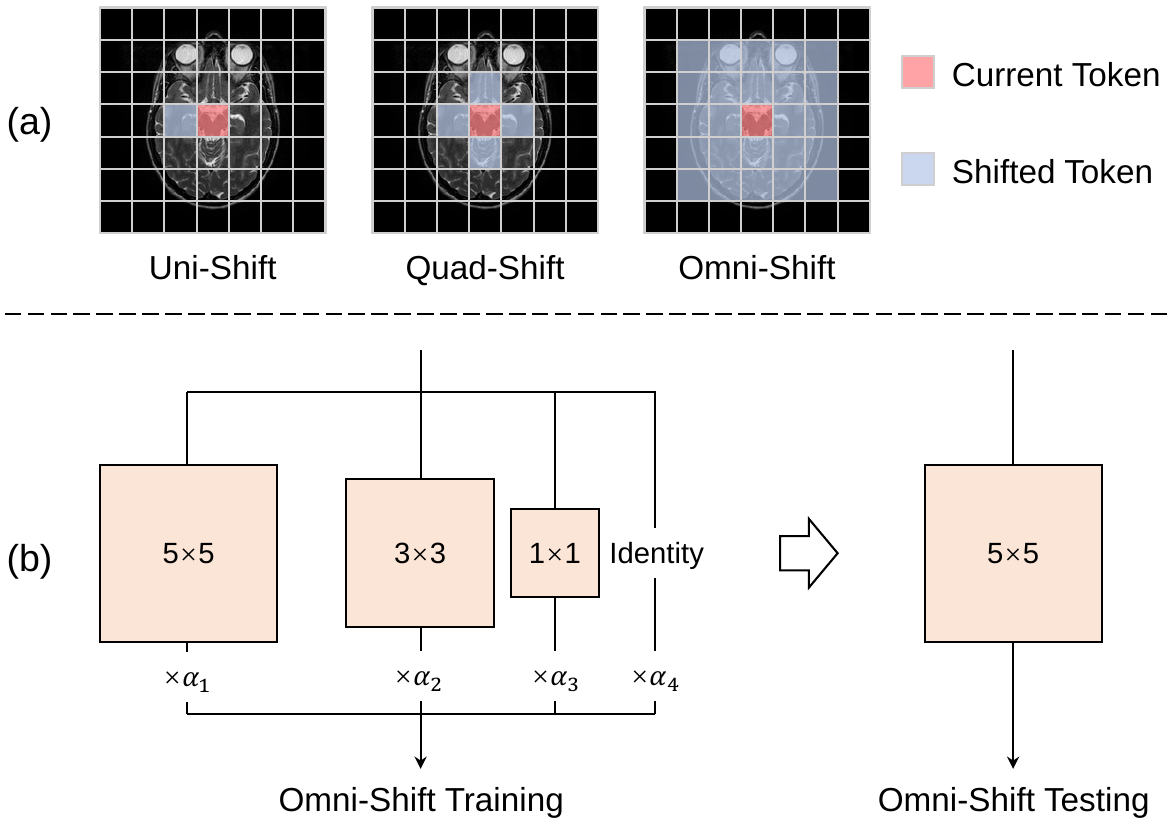}
\caption{(a) Illustrations of different token shift mechanisms. The Uni-Shift \cite{peng2023rwkv} fuses the current token with only the last (left) one by linear interpolation. The Quad-Shift \cite{duan2024vrwkv} fuses the current token with four adjacent tokens by linear interpolation. Our proposed Omni-Shift fuses the current token with tokens from all directions by convolution. (b) Illustration of the Omni-Shift with structural re-parameterization.}
\label{fig_onmi_shift}
\end{figure}

The token shift mechanism in RWKV \cite{peng2023rwkv} is proposed to capture the local context in a token sequence, where neighboring tokens are assumed to be correlated and share similar context information. It works by shifting neighboring tokens to fuse with individual tokens through simple linear interpolation \cite{peng2023rwkv,duan2024vrwkv}. However, as shown in Fig.~\ref{fig_onmi_shift} (a), existing token shift mechanisms, such as the unidirectional token shift (Uni-Shift) in RWKV \cite{peng2023rwkv} and the quad-directional token shift (Quad-Shift) in Vision-RWKV \cite{duan2024vrwkv}, only shift tokens from limited directions and do not fully exploit the spatial relationships inherent in 2D images, where neighboring tokens in all directions are correlated. To address this issue, we propose an omnidirectional token shift (Omni-Shift) mechanism that shifts and fuses neighboring tokens from all directions using convolution. To achieve an accurate and efficient token shift mechanism for aggregating local context, Omni-Shift employs a structural re-parameterization \cite{ding2021repvgg} over convolution, involving a multi-branch structure during the training phase and a single-branch structure during testing. We will elaborate on these details as follows.

\textbf{Multi-branch Training of Omni-Shift}. Considering the differences in importance across various context ranges, Omni-Shift employs a multi-branch structure during training (as shown in Fig.~\ref{fig_onmi_shift} (b)), with each branch responsible for shifting tokens within a specific context range. Given an input feature $X \in \mathbb{R}^{H \times W \times C}$, the mechanism of Omni-Shift can be formulated as follows:
\begin{equation}
\begin{aligned}
\operatorname{Omni-Shift}(X)= & \alpha_1\operatorname{DConv}_{5 \times 5}(X)+\alpha_2\operatorname{DConv}_{3 \times 3}(X) \\
& +\alpha_3\operatorname{DConv}_{1 \times 1}(X)+\alpha_4 X,
\end{aligned}
\end{equation}
where $\alpha_i$ denotes a learnable parameter for scaling the specific branch. $\operatorname{DConv}_{k \times k}(\cdot)$ denotes the efficient depth-wise convolution with kernel size of $k$. The final result of Omni-Shift is a fusion of four branches, with each branch specializing in a specific context range, resulting in an accurate token shift. 

\textbf{Single-branch Testing of Omni-Shift}. A multi-branch design inevitably introduces additional parameters and computational costs compared to a single branch. To mitigate this issue, Omni-Shift adopts the approach used in RepVGG \cite{ding2021repvgg}, which involves performing structural re-parameterization during testing. Given that the convolution weights of smaller kernels can be integrated with those of larger kernels by zero-padding the weights of the smaller kernels, the multi-branch design of Omni-Shift can be consolidated into a single branch using a convolution kernel size of 5 during testing (as shown in Fig.~\ref{fig_onmi_shift} (b)). This re-parameterization strategy guarantees both accuracy and testing-time efficiency of Omni-Shift.

\section{Experimental Setup}
\subsection{Dataset}
We conduct extensive experiments on various MedIR tasks to demonstrate the performance of our proposed Restore-RWKV model. These include three individual tasks—PET image synthesis, CT image denoising, and MRI image super-resolution—as well as an all-in-one medical image restoration task that combines the aforementioned three tasks for multi-task learning. For each individual MedIR task, we prepare two types of datasets: a base dataset for model training and in-distribution testing, and another dataset for out-of-distribution testing. The datasets corresponding to each task are described below.

\textbf{PET Image Synthesis.} \textbf{(1) PolarStar m660 Dataset}. The PolarStar m660 dataset is a private dataset utilized for both model training and in-distribution testing. It contains HQ PET images collected from 159 patients using the PolarStar m660 PET/CT system in list mode, with an average injection dose of 293MBq $^{18}$F-FDG. These 159 HQ PET images were divided into 120 for training, 10 for validation, and 29 for testing. LQ PET images are generated through random list mode decimation with a dose reduction factor (DRF) of 12. Both HQ and LQ PET images are reconstructed using the standard OSEM \cite{hudson1994osem} method. Each PET image has 3D shapes of 400$\times$192$\times$192 and is divided into 192 2D slices sized 400$\times$192. Slices containing only air are excluded. Consequently, the PolarStar m660 dataset contains 8350 2D PET images for training, 684 images for validation, and 2044 images for testing. \textbf{(2) PolarStar Flight Dataset}. The PolarStar Flight dataset is another private dataset for out-of-distribution testing. It contains HQ PET images collected from 10 patients using the PolarStar Flight PET/CT system in list mode, with an average injection dose of 293MBq $^{18}$F-FDG. The DRF for the LQ images is set to 4. The data preparation process is identical to that of the PolarStar m660 dataset. In total, the PolarStar Flight dataset contains 674 image pairs, each with a size of 400$\times$192.

\textbf{CT Image Denoising.}  \textbf{(1) AAPM Dataset.} The publicly available AAPM dataset \cite{mccollough2017AAPM} is employed for both model training and in-distribution testing. It comprises paired standard-dose HQ CT images and quarter-dose LQ CT images of the abdomen. These images are collected from 10 patients and are divided into 8 patients for training, 1 for validation, and 1 for testing. Each 3D CT image has a dimension of 512$\times$512$\times n$. Consequently, we extract 2D slices sized 512$\times$512 and obtain 2039 images for training, 128 for validation, and 211 for testing. \textbf{(2) LDCT Dataset.} The publicly available LDCT dataset \cite{moen2021LDCT_Dataset} is used for out-of-distribution testing. We randomly select 5 paired standard-dose HQ CT images and quarter-dose LQ CT images of the abdomen. After extracting 2D slices, the LDCT dataset contains 747 image pairs, each with dimensions of 512$\times$512.

\textbf{MRI Image Super-Resolution.} \textbf{(1) IXI Dataset.} The publicly available IXI dataset (\href{http://brain-development.org/ixi-dataset/}{http://brain-development.org/ixi-dataset/}) is exploited for both model training and in-distribution testing. The IXI dataset consists of 578 HQ T2-weighted 3D brain MRI images. These images are divided into 405 for training, 59 for validation, and 114 for testing. Each 3D MRI image has dimensions of 256$\times$256$\times n$, from which we extract the central 100 2D slices sized 256$\times$256 to exclude side slices. This results in 40500 high-quality MRI slices for training, 5828 for validation, and 11400 for testing. Following prior work \cite{zhao2019channel,lei2023mcvar}, the LQ image is generated by transforming the HQ image to the frequency domain, retaining only the central 6.25$\%$ of frequency data points while zero-filling the high-frequency parts, and then converting it back to the image domain. \textbf{(2) HCP Dataset.} The publicly available HCP dataset \cite{van2013hcp} is exploited for out-of-distribution testing. A total of 50 HQ T2-weighted 3D brain MRI images are randomly selected as HQ images. The procedure for generating LQ images is identical to that used for the IXI dataset. From each 3D image, the central 100 slices are extracted, resulting in 5,000 LQ-HQ image pairs, each with dimensions of 320$\times$256.

\begin{table*}[t] 
\caption{PET image synthesis results. The best result is marked in bold, and the second-best result is underlined.} 
\centering 
\resizebox{\textwidth}{!}{
\begin{tabular}{ccccccclclclclclc}
\hline
                         &  &                              &  &                             &  & \multicolumn{5}{c}{PolarStar m660}                                 &  & \multicolumn{5}{c}{PolarStar Flight}                               \\ \cline{7-11} \cline{13-17} 
\multirow{-2}{*}{Method} &  & \multirow{-2}{*}{Params (M)} &  & \multirow{-2}{*}{FLOPs (G)} &  & PSNR↑          &  & SSIM↑           &  & RMSE↓           &  & PSNR↑          &  & SSIM↑           &  & RMSE↓           \\ \hline
Xiang's                  &  & 0.23                         &  & 3.77                        &  & 35.93          &  & 0.9167          &  & 0.0980          &  & 33.85          &  & 0.9092          &  & 0.1320          \\
Xu's                     &  & 0.52                         &  & 1.11                        &  & 36.44          &  & 0.9376          &  & 0.0982          &  & 34.14          &  & 0.9160          &  & 0.1275          \\
DCNN                     &  & 1.33                         &  & 21.86                       &  & 36.27          &  & 0.9243          &  & 0.0954          &  & 33.96          &  & 0.9135          &  & 0.1303          \\
CycleWGAN                &  & 1.00                         &  & 16.42                       &  & 36.62          &  & 0.9290          &  & 0.0910          &  & 34.12          &  & 0.9132          &  & 0.1279          \\
DCITN                    &  & 0.08                         &  & 1.24                        &  & 36.09          &  & 0.9285          &  & 0.0970          &  & 33.68          &  & 0.9064          &  & 0.1346          \\
ARGAN                    &  & 31.14                        &  & 8.48                        &  & 36.73          &  & 0.9406          &  & 0.0902          &  & 34.13          &  & 0.9191          &  & 0.1279          \\
DRMC                     &  & 0.62                         &  & 9.92                        &  & 36.00          &  & 0.9352          &  & 0.0998          &  & 33.62          &  & 0.9081          &  & 0.1356          \\
\rowcolor[HTML]{EFEFEF} 
Restore-RWKV-light       &  & 1.16                         &  & 1.52                        &  & {\ul 36.96}    &  & {\ul 0.9427}    &  & {\ul 0.0887}    &  & {\ul 34.44}    &  & {\ul 0.9223}    &  & {\ul 0.1232}    \\
\rowcolor[HTML]{EFEFEF} 
Restore-RWKV             &  & 27.91                        &  & 37.46                       &  & \textbf{37.33} &  & \textbf{0.9474} &  & \textbf{0.0852} &  & \textbf{34.70} &  & \textbf{0.9256} &  & \textbf{0.1212} \\ \hline
\end{tabular}
}
\label{tab_pet_synthesis}
\end{table*}

\begin{figure*}[t]
\centering
\includegraphics[width=\textwidth]{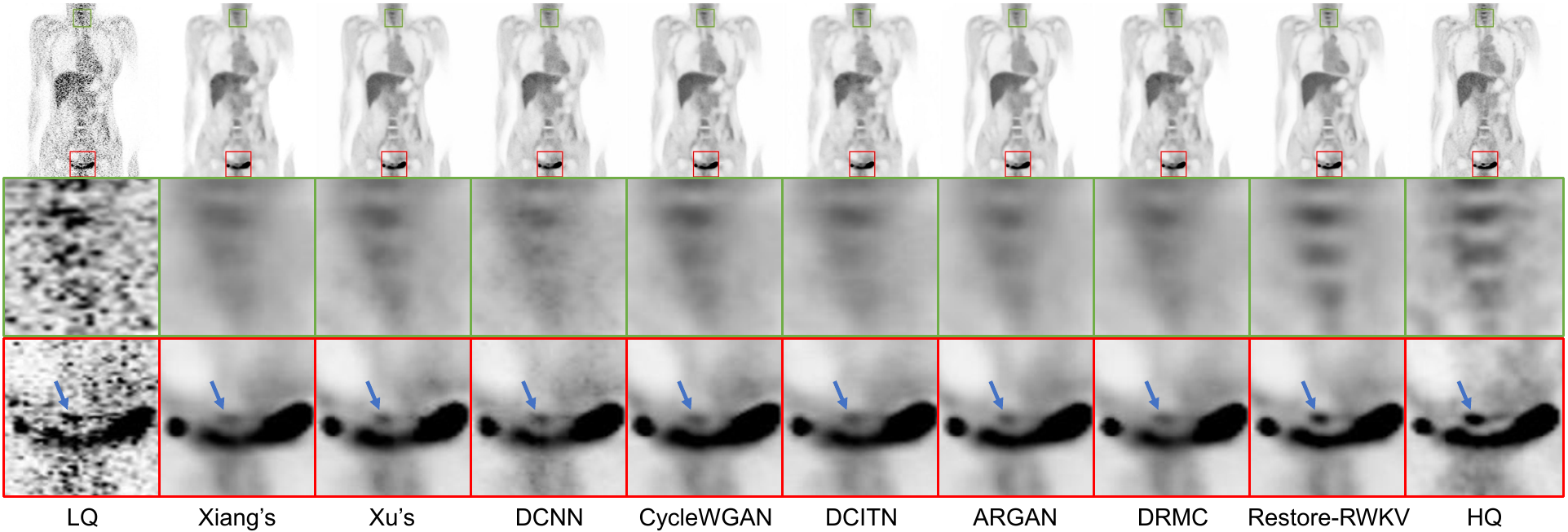}
\caption{Visual comparison of different methods in PET image synthesis. The zoomed-in rectangular region is recommended for a better comparison.}
\label{fig_pet_comparison_vis}
\end{figure*}

\begin{figure}[htb]
\centering
\includegraphics[width=0.48\textwidth]{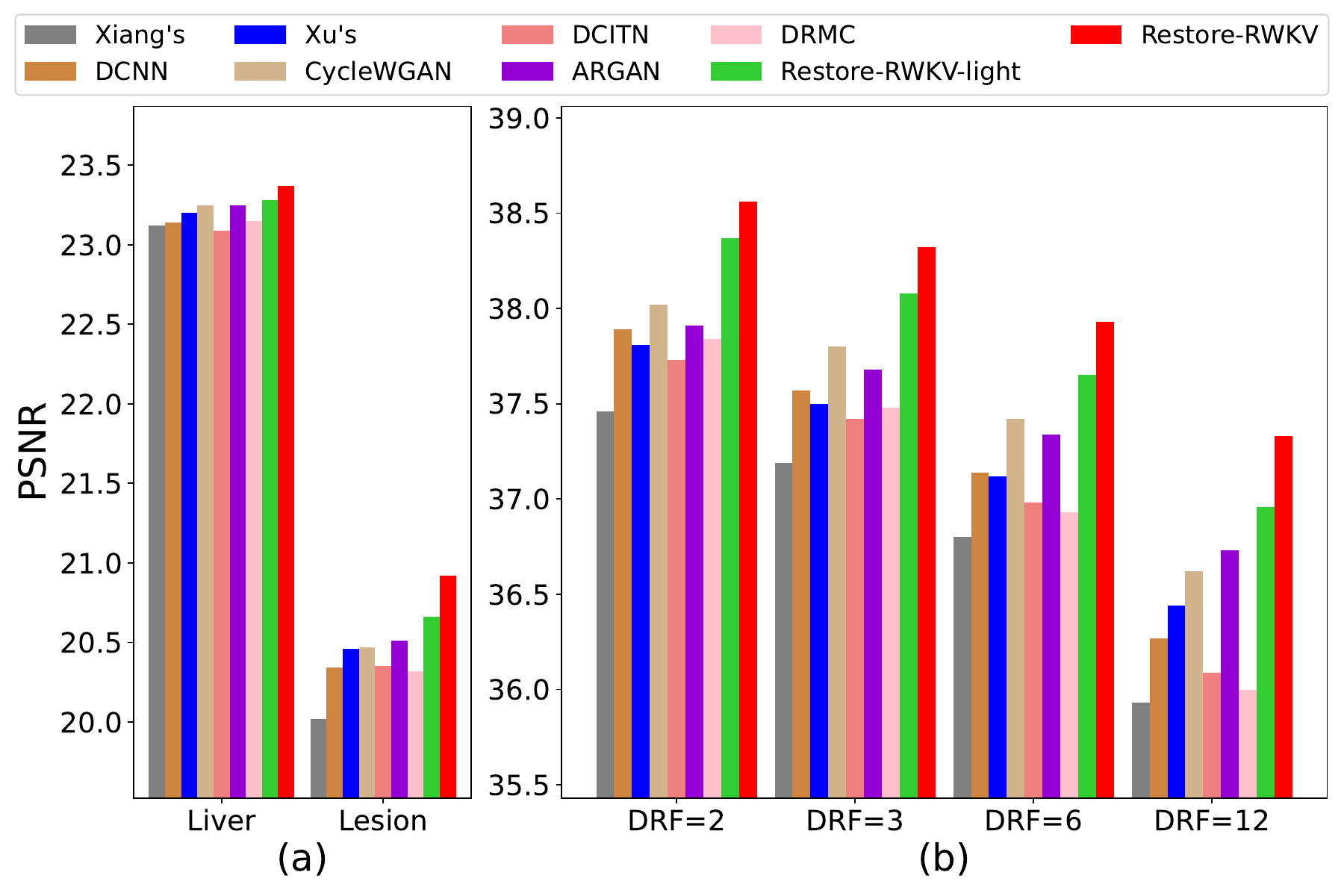}
\caption{Performance comparison on (a) clinical ROIs and (b) low-dose PET images with different DRFs.}
\label{fig_psnr_hist}
\end{figure}

\textbf{All-in-One Medical Image Restoration.} The objective of all-in-one medical image restoration \cite{yang2024amir} is to train a universal model capable of handling multiple MedIR tasks. Therefore, the dataset used in this study combines datasets from the aforementioned three individual tasks: the PolarStar m660 dataset for PET image synthesis, the AAPM dataset for CT image denoising, and the IXI dataset for MRI image super-resolution. The training, validation, and testing splits for each dataset remain consistent with their original partitions used in the individual tasks, but they are combined into a unified dataset for this study.

\subsection{Implementation} 
For the architecture of Restore-RWKV, the number of R-RWKV blocks is set to $N_1=N_{\text{refinement}}=4$, $N_2=N_3=6$, and $N_4=8$. The number of input channels is $C=48$, and the number of attention recurrences in Re-WKV is $M=2$. We also introduce two additional model variants: the lightweight variant, Restore-RWKV-light, and the all-in-one medical image restoration variant, Restore-RWKV-routing. The Restore-RWKV-light differs from the standard Restore-RWKV in both the number of blocks and channels, with values set to $N_1=N_2=N_4=N_{\text{refinement}}=1$, $N_3=4$ and $C=16$. The Restore-RWKV-routing variant adopts the task-adaptive routing strategy introduced by the AMIR model \cite{yang2024amir}, integrating a spatial routing module at the beginning of each encoder and decoder level in Restore-RWKV. To maintain parameter efficiency, the channel number is reduced to $C=40$

For model training, we use training patches of size 128$\times$128 and a batch size of 4. All model variants of Restore-RWKV are trained using the Adam optimizer and $L_1$ loss for 30K iterations. The learning rate is initialized at 2$e^{-4}$ and gradually reduced to 1$e^{-6}$ using cosine annealing. All experiments are conducted in PyTorch, utilizing an NVIDIA A100 GPU.

\subsection{Evaluation} 
To evaluate image quality, we use three widely recognized metrics for quantitative comparisons: Peak Signal-to-Noise Ratio (PSNR), Structural Similarity Index (SSIM), and Root Mean Squared Error (RMSE). Higher PSNR and SSIM values indicate better performance, while lower RMSE values signify improved results. To further assess model performance on clinically relevant regions of interest (ROIs), a radiologist annotates the liver and lesion in the testing dataset of PolarStar m660 for PET image synthesis. PSNR is then computed specifically over these ROIs. The liver is selected because it is commonly used to evaluate the uniformity and noise levels of PET images, while the lesion is prioritized as it represents the most critical region for disease diagnosis. To evaluate model efficiency, we calculate the number of model parameters and floating-point operations per second (FLOPs). The input image size is set to 128$\times$128 when calculating FLOPs.

\section{Experimental Results} 
\subsection{PET Image Synthesis Results} 
For PET image synthesis, we compare Restore-RWKV with seven PET image synthesis methods: Xiang's model \cite{xiang2017xiang}, DCNN \cite{chan2018dcnn}, Xu's model \cite{xu2017200x}, CycleWGAN \cite{zhou2020cyclewgan}, DCITN \cite{zhou2022sgsgan}, ARGAN \cite{luo2022argan} and DRMC \cite{yang2023drmc}. Table \ref{tab_pet_synthesis} indicates that Restore-RWKV significantly (p-value $<$ 0.05 in the t-test) outperforms other methods on both the in-distribution PolarStar m660 dataset and out-of-distribution PolarStar Flight dataset. This demonstrates the effectiveness and strong model generalization ability of the proposed Restore-RWKV. Notably, even the lightweight variant Restore-RWKV-light can already outperform those comparison methods. Fig.~\ref{fig_pet_comparison_vis} presents the visual comparison results. It is clear that the comparison methods generally fail to differentiate between noise and image details, resulting in the smoothing of important features during denoising. In contrast, the proposed Restore-RWKV is more effective at preserving these details, such as the sharpness of spine contours and the contrast in high-uptake regions. This advantage stems from Restore-RWKV's global receptive field, which enables it to leverage both local and global contexts to better distinguish between noise and image details, leading to superior recovery of fine details. 

To further evaluate the model's effectiveness and generalization ability, we assess its performance on clinical ROIs and low-dose PET images with varying DRFs. Figure~\ref{fig_psnr_hist} (a) shows the model's performance on clinical ROIs of the liver and lesion. The proposed Restore-RWKV method most effectively preserves these two regions, especially the lesion region, where it outperforms all comparison methods by a large margin. Figure~\ref{fig_psnr_hist} (b) illustrates the model's performance on low-dose PET images with different DRFs. It is evident that Restore-RWKV consistently performs the best across all DRF levels, demonstrating its strong generalization ability.

\begin{table*}[!htb]
\caption{CT image denoising results. The best result is marked in bold, and the second-best result is underlined.}
\centering
\resizebox{\textwidth}{!}{
\begin{tabular}{ccccccccccccccclc}
\hline
                         &  &                              &  &                             &  & \multicolumn{5}{c}{AAPM}                                                   &           & \multicolumn{5}{c}{LDCT}                                          \\ \cline{7-11} \cline{13-17} 
\multirow{-2}{*}{Method} &  & \multirow{-2}{*}{Params (M)} &  & \multirow{-2}{*}{FLOPs (G)} &  & PSNR↑          &           & SSIM↑           &           & RMSE↓           &           & PSNR↑          &           & SSIM↑           &  & RMSE↓           \\ \hline
REDCNN                   &  & 1.85                         &  & 24.92                       &  & 33.19          &           & 0.9113          &           & 8.9427          &           & 35.07          &           & 0.9201          &  & 7.2062          \\
WGAN-VGG                 &  & 0.06                         &  & 0.92                        &  & 31.96          &           & 0.8977          &           & 10.3366         &           & 33.72          &           & 0.9004          &  & 8.4734          \\
EDCNN                    &  & 0.08                         &  & 1.32                        &  & 33.41          &           & 0.9155          &           & 8.7401          &           & 35.37          &           & 0.9242          &  & 6.9923          \\
TransCT                  &  & 13.23                        &  & 0.65                        &  & 32.62          &           & 0.9082          &           & 9.5330          &           & 34.64          &           & 0.9183          &  & 7.5752          \\
Eformer                  &  & 0.34                         &  & 1.28                        &  & 33.35          &           & 0.9175          &           & 8.8030          &           & 35.48          &           & 0.9262          &  & 6.9151          \\
CTformer                 &  & 1.45                         &  & 3.44                        &  & 33.25          &           & 0.9134          &           & 8.8974          &           & 35.31          &           & 0.9240          &  & 7.0475          \\
DenoMamba                &  & 112.62                       &  & 110.24                      &  & 33.53          &           & 0.9149          &           & 8.6115          &           & 35.67          &           & 0.9259          &  & 6.7565          \\
\rowcolor[HTML]{EFEFEF} 
Restore-RWKV-light       &  & 1.16                         &  & 1.52                        &  & {\ul 33.64}    & {\ul }    & {\ul 0.9178}    & {\ul }    & {\ul 8.5141}    & {\ul }    & {\ul 35.86}    & {\ul }    & {\ul 0.9286}    &  & {\ul 6.6108}    \\
\rowcolor[HTML]{EFEFEF} 
Restore-RWKV             &  & 27.91                        &  & 37.46                       &  & \textbf{33.80} & \textbf{} & \textbf{0.9198} & \textbf{} & \textbf{8.3600} & \textbf{} & \textbf{36.06} & \textbf{} & \textbf{0.9310} &  & \textbf{6.4614} \\ \hline
\end{tabular}
}
\label{tab_ct_denoising}
\end{table*} 

\begin{figure*}[!t]
\centering
\includegraphics[width=\textwidth]{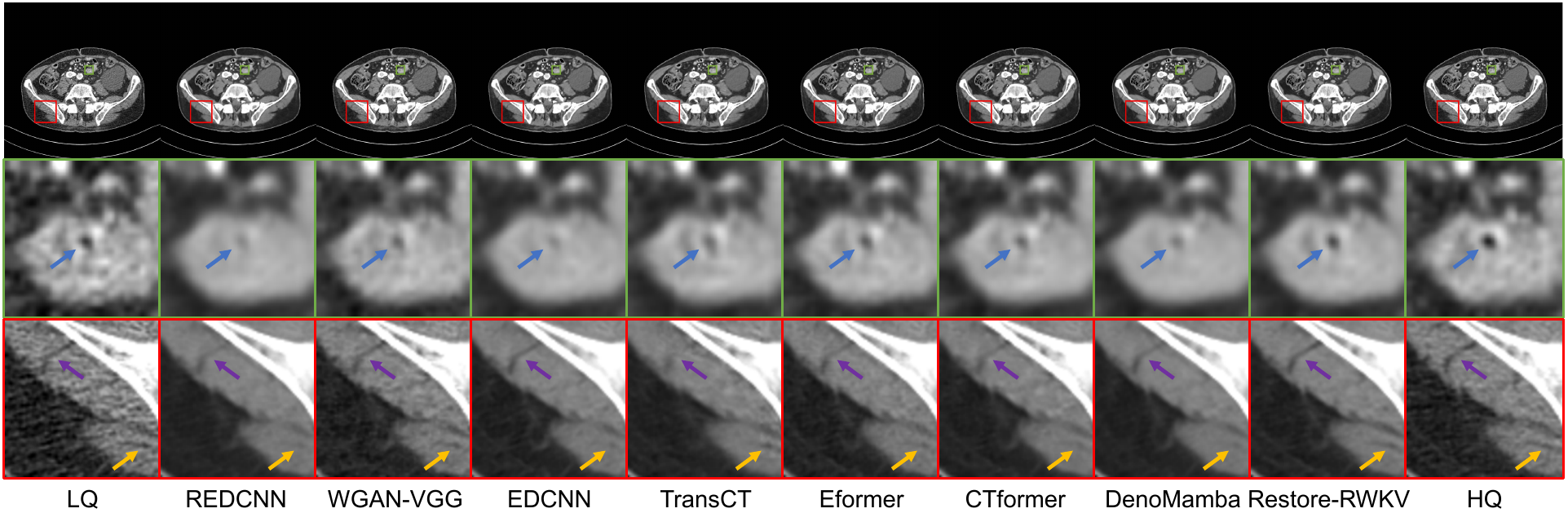}
\caption{Visual comparison of different methods in CT image denoising. The zoomed-in rectangular region is recommended for a better comparison.}
\label{fig_ct_comparison_vis}
\end{figure*}

\subsection{CT Image Denoising Results} 
For CT image denoising, we compare our proposed Restore-RWKV with seven CT image denoising methods: REDCNN \cite{chen2017redcnn}, WGAN-VGG \cite{yang2018wgan_vgg}, EDCNN \cite{liang2020edcnn}, TransCT \cite{zhang2021transct}, Eformer \cite{luthra2021eformer}, CTformer \cite{wang2023ctformer},  and DenoMamba \cite{ozturk2024denomamba}. Table \ref{tab_ct_denoising} indicates that Restore-RWKV achieves the best CT image denoising performance (p-value $<$ 0.05 in the t-test) compared to all other methods on both the in-distribution dataset AAPM and the out-of-distribution dataset LDCT. Even the Restore-RWKV-light, with only 1.16 million parameters, outperforms the best-performing comparison method, DenoMamba, which has 112.62 million parameters. These results indicate that Restore-RWKV is an efficient and effective model for CT denoising. The visual comparison result in Fig.~\ref{fig_ct_comparison_vis} (a) shows that the image recovered by Restore-RWKV is sharper and clearer.

\begin{table*}[!htb]
\caption{MRI image super-resolution results. The best result is marked in bold, and the second-best result is underlined.}
\centering 
\resizebox{\textwidth}{!}{
\begin{tabular}{ccccccccccccccccc}
\hline
                         &  &                              &  &                             &  & \multicolumn{5}{c}{IXI}                                                     &           & \multicolumn{5}{c}{HCP}                                                     \\ \cline{7-11} \cline{13-17} 
\multirow{-2}{*}{Method} &  & \multirow{-2}{*}{Params (M)} &  & \multirow{-2}{*}{FLOPs (G)} &  & PSNR↑          &           & SSIM↑           &           & RMSE↓            &           & PSNR↑          &           & SSIM↑           &           & RMSE↓            \\ \hline
VDSR                     &  & 0.66                         &  & 10.89                       &  & 30.04          &           & 0.9140          &           & 36.0508          &           & 27.74          &           & 0.8191          &           & 77.1382          \\
DAGAN                    &  & 98.59                        &  & 33.98                       &  & 30.55          &           & 0.9189          &           & 34.0866          &           & 28.90          &           & 0.8630          &           & 67.5468          \\
FAWDN                    &  & 5.07                         &  & 82.92                       &  & 30.04          &           & 0.9136          &           & 35.9564          &           & 28.97          &           & 0.8665          &           & 67.0522          \\
SwinMR                   &  & 11.40                        &  & 200.18                      &  & 30.93          &           & 0.9253          &           & 32.7339          &           & 29.03          &           & 0.8666          &           & 66.5876          \\
SDAUT                    &  & 67.23                        &  & 55.25                       &  & 30.96          &           & 0.9257          &           & 32.5928          &           & 29.08          &           & 0.8710          &           & 66.2161          \\
F-UNet                   &  & 32.12                        &  & 10.21                       &  & 31.26          &           & 0.9314          &           & 31.5675          &           & 29.08          &           & 0.8697          &           & 66.1355          \\
MambaIR                  &  & 31.50                        &  & 34.35                       &  & {\ul 31.77}    & {\ul }    & {\ul 0.9369}    & {\ul }    & {\ul 29.8372}    & {\ul }    & {\ul 29.29}    & {\ul }    & {\ul 0.8751}    & {\ul }    & {\ul 64.6289}    \\
\rowcolor[HTML]{EFEFEF} 
Restore-RWKV-light       &  & 1.16                         &  & 1.52                        &  & 31.36          &           & 0.9309          &           & 31.2564          &           & 29.22          &           & 0.8738          &           & 65.1056          \\
\rowcolor[HTML]{EFEFEF} 
Restore-RWKV             &  & 27.91                        &  & 37.46                       &  & \textbf{32.09} & \textbf{} & \textbf{0.9408} & \textbf{} & \textbf{28.9713} & \textbf{} & \textbf{29.41} & \textbf{} & \textbf{0.8788} & \textbf{} & \textbf{63.8063} \\ \hline
\end{tabular}
}
\label{tab_mri_sr}
\end{table*}

\begin{figure*}[!htb]
\centering
\includegraphics[width=\textwidth]{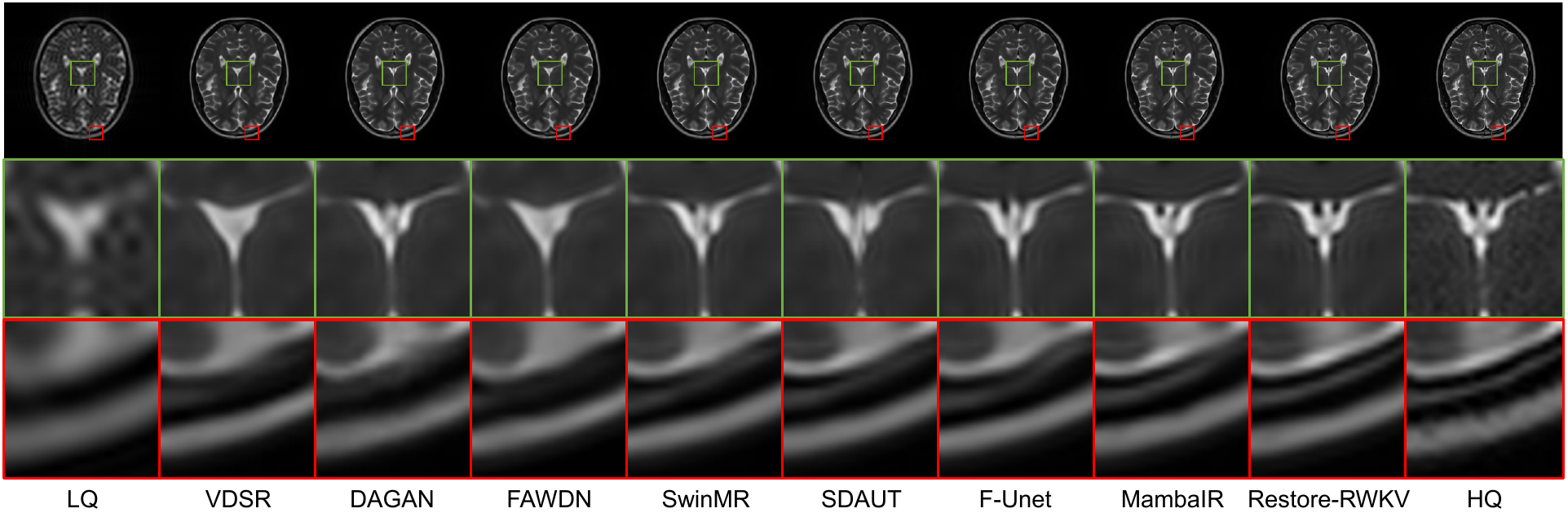}
\caption{Visual comparison of different methods in MRI image super-resolution. The zoomed-in rectangular region is recommended for a better comparison}
\label{fig_mri_comparison_vis}
\end{figure*}

\begin{table*}[!htb]
\caption{All-in-one medical image restoration results. The best result is marked in bold, and the second-best result is underlined.}
\centering
\resizebox{\textwidth}{!}{
\begin{tabular}{clclclclclclclclclclclc}
\hline
                         &                                              &                              &                                              &                             &                                              & \multicolumn{5}{c}{PET Image Synthesis}                  &  & \multicolumn{5}{c}{CT Image Denoising}                   &  & \multicolumn{5}{c}{MRI Image   Super-Resolution}                                                      \\ \cline{7-11} \cline{13-17} \cline{19-23} 
\multirow{-2}{*}{Method} &                                              & \multirow{-2}{*}{Params (M)} &                                              & \multirow{-2}{*}{FLOPs (G)} &                                              & PSNR↑          &  & SSIM↑           &  & RMSE↓           &  & PSNR↑          &  & SSIM↑           &  & RMSE↓           &  & PSNR↑          &                                              & SSIM↑           &  & RMSE↓            \\ \hline
ARGAN                    &                                              & 31.14                        &                                              & 8.48                        &                                              & 36.75          &  & 0.9389          &  & 0.0907          &  & 32.92          &  & 0.9111          &  & 9.2110          &  & 30.08          &                                              & 0.9083          &  & 35.7999          \\
DenoMamba                &                                              & 112.62                       &                                              & 110.24                      &                                              & 36.81          &  & 0.9367          &  & 0.0895          &  & 33.18          &  & 0.9115          &  & 8.9512          &  & 30.32          &                                              & 0.9091          &  & 34.6972          \\
MambaIR                  &                                              & 31.50                        &                                              & 34.35                       &                                              & 37.17          &  & 0.9458          &  & {\ul 0.0864}    &  & 33.50          &  & 0.9165          &  & 8.6345          &  & 31.31          &                                              & 0.9305          &  & 31.3150          \\
\rowcolor[HTML]{EFEFEF} 
Restore-RWKV             &                                              & 27.98                        &                                              & 37.46                       &                                              & {\ul 37.21}    &  & {\ul 0.9475}    &  & 0.0865          &  & 33.69          &  & {\ul 0.9188}    &  & 8.4609          &  & 31.94          &                                              & 0.9386          &  & 29.3562          \\ \hline
AirNet                   &                                              & 7.61                         &                                              & 230.48                      &                                              & 37.17          &  & 0.9451          &  & {\ul 0.0864}    &  & 33.62          &  & 0.9176          &  & 8.5226          &  & 31.39          &                                              & 0.9316          &  & 31.1141          \\
DRMC                     &                                              & 0.62                         &                                              & 9.92                        &                                              & 36.19          &  & 0.9376          &  & 0.0960          &  & 33.28          &  & 0.9153          &  & 8.8674          &  & 29.55          &                                              & 0.9032          &  & 38.1691          \\
AMIR                     &                                              & 23.54                        &                                              & 31.76                       &                                              & 37.12          &  & {\ul 0.9475}    &  & 0.0876          &  & {\ul 33.70}    &  & 0.9182          &  & {\ul 8.4520}    &  & {\ul 32.03}    &                                              & {\ul 0.9396}    &  & {\ul 29.0988}    \\
\rowcolor[HTML]{EFEFEF} 
Restore-RWKV-routing     & \multicolumn{1}{c}{\cellcolor[HTML]{EFEFEF}} & 24.85                        & \multicolumn{1}{c}{\cellcolor[HTML]{EFEFEF}} & 31.95                       & \multicolumn{1}{c}{\cellcolor[HTML]{EFEFEF}} & \textbf{37.32} &  & \textbf{0.9479} &  & \textbf{0.0852} &  & \textbf{33.74} &  & \textbf{0.9189} &  & \textbf{8.4148} &  & \textbf{32.10} & \multicolumn{1}{c}{\cellcolor[HTML]{EFEFEF}} & \textbf{0.9408} &  & \textbf{28.9507} \\ \hline
\end{tabular}
}
\label{tab_all_in_one}
\end{table*}

\subsection{MRI Image Super-Resolution Results}
For MRI image super-resolution, we compare our proposed Restore-RWKV with seven image super-resolution methods: VDSR \cite{kim2016vdsr}, DAGAN \cite{yang2017dagan}, FAWDN \cite{chen2020fawdn}, SwinMR \cite{huang2022swinmr}, SDAUT \cite{huang2022sdaut}, F-Unet \cite{sun2025funet}, and MambaIR \cite{guo2025mambair}. Table \ref{tab_mri_sr} demonstrates that Restore-RWKV significantly (p-value $<$ 0.05 in the t-test) outperforms all comparison methods on both the in-distribution dataset IXI and the out-of-distribution dataset HCP in MRI image super-resolution. The Mamba-based model, MambaIR, which is also reported to achieve a global receptive field with linear complexity, achieves the second-best result. However, as illustrated in Fig.\ref{fig_erf}, Restore-RWKV can achieve a more significant global receptive field than MambaIR and, as a result, better captures global dependencies to recover image structures and details. Fig.\ref{fig_mri_comparison_vis} presents visualization results across different methods, highlighting that Restore-RWKV excels in recovering realistic MRI image details.

\subsection{All-in-One Medical Image Restoration Results} 
The aforementioned experimental results indicate that Restore-RWKV achieves superior performance in individual MedIR tasks. To further evaluate the model's capacity and generalization ability, we assess its performance on the all-in-one medical image restoration task, which involves using a single model to handle multiple MedIR tasks. Given the apparent disparities among different MedIR tasks, this all-in-one task is particularly challenging and serves as a robust test of model capacity and generalization ability. For a fair comparison with all-in-one models, we also introduce a model variant Restore-RWKV-routing, which incorporates the task-adaptive routing strategy developed by the AMIR model \cite{yang2024amir}. We compare our proposed Restore-RWKV and Restore-RWKV-routing with three best-performing comparison methods in each individual task: ARGAN \cite{luo2022argan}, DenoMamba \cite{ozturk2024denomamba}, and MambaIR \cite{guo2025mambair}, as well as three all-in-one image restoration methods: AirNet \cite{li2022airnet}, DRMC \cite{yang2023drmc}, and AMIR \cite{yang2024amir}.  As shown in Table \ref{tab_all_in_one}, Restore-RWKV outperforms the three best-performing comparison methods from individual tasks, indicating that it has the best model capacity and generalization ability to deal with different tasks. Meanwhile, Restore-RWKV-routing achieves the best overall results, surpassing all other all-in-one methods. This suggests that Restore-RWKV is a robust restoration backbone that can be easily extended to multi-task settings through the integration of specific modules.

\subsection{Ablation Studies} 
We conduct ablation experiments on the IXI dataset for MRI image super-resolution task to investigate the significance of two key innovations in the Restore-RWKV: Re-WKV attention and Omni-Shift mechanisms.

\begin{figure}[!t] 

\centering
\includegraphics[width=0.48\textwidth]{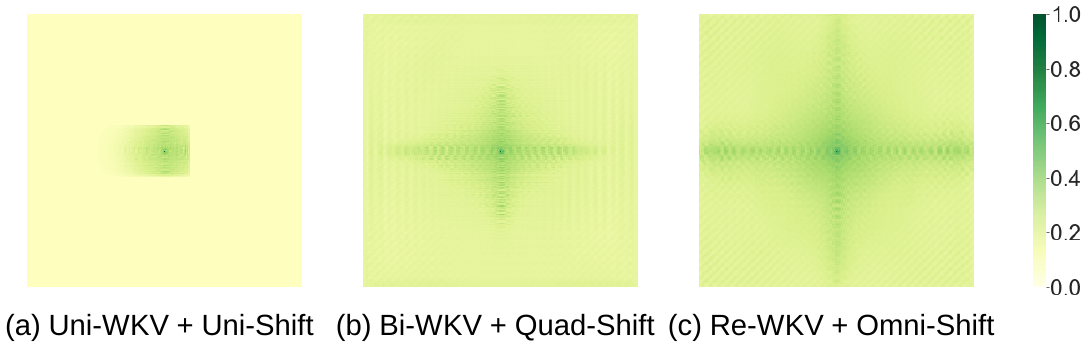}
\caption{Comparison of the effective receptive field (ERF) in models with various WKV attention and token shift combinations. (a) The combination of "Uni-WKV + Uni-Shift" in the original RWKV \cite{peng2023rwkv}. (b) The combination of "Bi-WKV + Quad-Shift" in the Vision-RWKV \cite{duan2024vrwkv}. (c) The combination of "Re-WKV + Omni-Shift" in the proposed Restore-RWKV. A more extensively distributed dark area indicates a larger ERF.}
\label{fig_erf_ablation}
\end{figure}

\textbf{Effect of Re-WKV and Omni-Shift.} To evaluate the effectiveness of Re-WKV attention and Omni-Shift, we conduct experiments by replacing them with other WKV attention layers, such as Uni-WKV \cite{peng2023rwkv} and Bi-WKV \cite{duan2024vrwkv}, as well as token shift layers, including Uni-Shift \cite{peng2023rwkv} and Quad-Shift \cite{duan2024vrwkv}. Table \ref{tab_component_analysis} indicates that both Re-WKV and Omni-Shift can effectively improve model performance, with the combination of both achieving the best results. Notably, compared to the original RWKV's \cite{peng2023rwkv} combination of "Uni-WKV + Uni-Shift," our proposed combination "Re-WKV + Omni-Shift" achieves an improvement of over 0.77 dB in PSNR. Fig.~\ref{fig_erf_ablation} illustrates the impact of different attention and token shift combinations on the effective receptive field (ERF) of models. Generally, a larger receptive field allows the model to capture information from a wider region, thereby enhancing restoration performance. We can see that the original RWKV's \cite{peng2023rwkv} combination of "Uni-WKV + Uni-Shift" achieves only a local receptive field, while the Vision-RWKV's \cite{duan2024vrwkv} combination of "Bi-WKV + Quad-Shift" achieves a global receptive field. Our proposed Restore-RWKV's combination of "Re-WKV + Omni-Shift" achieves the largest global receptive field. This indicates that the combination of "Re-WKV + Omni-Shift" most effectively ensures the model's ability to capture global dependencies in 2D images.

\textbf{Ablation on Re-WKV.} We investigate the influence of attention recurrence numbers on Re-WKV. As shown in Table \ref{tab_recurrence}, the model performance increases with the number of recurrences $M$ in Re-WKV attention. Especially, $M=2$ achieves a significant improvement over $M=1$. This is because conducting attention from two different scan directions ($M=2$) enables more effective token interaction on 2D images than using only a single direction ($M=1$). As further increasing the recurrence number does not yield a significant performance improvement, we finally adopt $M=2$ in Restore-RWKV.

We also explore architectural designs for conducting attention from both horizontal and vertical directions. We consider three designs. The first design alternates the scan direction between two adjacent blocks. The second design conducts attention from both directions within a block individually and then sums the attention results. The third design is our proposed recurrent attention, which recurrently conducts attention from both directions within a block. The results are shown in Table \ref{tab_attention_design} and our proposed design with recurrent attention achieves the best performance.

\textbf{Ablation on Omni-Shift.} We conduct experiments to investigate two factors that most influence the performance of Omni-Shift: the context range of token shift and the parameterization strategy. As shown in Table \ref{tab_omni_shift}, the experimental results indicate that Omni-Shift achieves the best performance with the context range set to $k=5$. Additionally, the reparameterization strategy significantly improves the accuracy of token shift, thereby enhancing MedIR performance. 

\begin{table}[!t]
\caption{Component analysis of Re-WKV and Omni-Shift. The \textbf{best} results are highlighted.}
\centering
\resizebox{0.48\textwidth}{!}{
\begin{tabular}{ccccccc}
\hline
Method                     &  & PSNR↑          &           & SSIM↑           &           & RMSE↓            \\ \hline
Uni-WKV + Uni-Shift        &  & 31.32          &           & 0.9315          &           & 31.3278          \\
Bi-WKV + Q-Shift           &  & 31.73          &           & 0.9359          &           & 30.1125          \\
Re-WKV + Q-Shift           &  & 31.93          &           & 0.9386          &           & 29.4302          \\
Bi-WKV + Omni-Shift        &  & 31.94          &           & 0.9384          &           & 29.4318          \\
Re-WKV + Omni-Shift (Ours) &  & \textbf{32.09} & \textbf{} & \textbf{0.9408} & \textbf{} & \textbf{28.9713} \\ \hline
\end{tabular}
}
\label{tab_component_analysis}
\end{table} 

\begin{table}[!t]
\caption{Ablation study on the recurrence number in Re-WKV. The \textbf{best} results are highlighted.}
\centering
\resizebox{0.48\textwidth}{!}{
\begin{tabular}{ccccccc}
\hline
Recurrence   Number &  & PSNR↑          &           & SSIM↑           &           & RMSE↓            \\ \hline
$M=1$ (H-Scan)      &  & 31.94          &           & 0.9384          &           & 29.4318          \\
$M=1$ (V-Scan)      &  & 31.93          &           & 0.9383          &           & 29.4371          \\
$M=2$ (Ours)        &  & 32.09          &           & 0.9408          &           & 28.9713          \\
$M=3$               &  & 32.11          &           & 0.9409 & \textbf{} & 28.8183          \\
$M=4$               &  & \textbf{32.11} & \textbf{} & \textbf{0.9409} & \textbf{} & \textbf{28.7891} \\ \hline
\end{tabular}
}
\label{tab_recurrence}
\end{table} 

\begin{table}[!t]
\caption{Ablation study on architecture designs for conducting attention from both horizontal and vertical scan directions. The \textbf{best} results are highlighted.}
\centering
\resizebox{0.48\textwidth}{!}{
\begin{tabular}{ccccccc}
\hline
Architecture                 &  & PSNR↑          &           & SSIM↑           &           & RMSE↓            \\ \hline
Alternating between Blocks   &  & 31.97          &           & 0.9393          &           & 29.2942          \\
Sum in a Block               &  & 31.90          &           & 0.9386          &           & 29.4702          \\
Recurrence in a Block (Ours) &  & \textbf{32.09} & \textbf{} & \textbf{0.9408} & \textbf{} & \textbf{28.9713} \\ \hline
\end{tabular}
}
\label{tab_attention_design}
\end{table}

\begin{table}[!t]
\caption{Ablation study on Omni-Shift. The \textbf{best} results are highlighted.}
\centering
\resizebox{0.48\textwidth}{!}{
\begin{tabular}{ccccccc}
\hline
Context   Range     &  & PSNR↑          &           & SSIM↑           &           & RMSE↓            \\ \hline
$k=1$               &  & 31.85          &           & 0.9375          &           & 29.7197          \\
$k=3$               &  & 31.99          &           & 0.9394          &           & 29.2586          \\
$k=5$ (Ours)        &  & \textbf{32.09} & \textbf{} & \textbf{0.9408} & \textbf{} & \textbf{28.9713} \\
$k=5$ (w/o reparam) &  & 32.01          &           & 0.9396 &  & 29.1992          \\
$k=7$               &  & 32.06 &  & 0.9396 & & 28.9602 \\
$k=9$               &  & 32.03          &           & 0.9398          &           & 29.1360          \\ \hline
\end{tabular}
}
\label{tab_omni_shift}
\end{table}

\section{Conclusion}
In this paper, we propose Restore-RWKV, leveraging the advanced RWKV architecture for the first time in the field of medical image restoration (MedIR). To achieve superior restoration performance, Restore-RWKV innovatively incorporates a recurrent WKV (Re-WKV) attention mechanism with linear computational complexity and an omnidirectional token shift (Omni-Shift) mechanism to effectively capture global and local dependencies in 2D images, respectively. These innovations make Restore-RWKV an efficient and effective backbone for medical image restoration. Notably, even a lightweight variant with only 1.16 million parameters outperforms or achieves comparable results to existing state-of-the-art methods. Extensive experiments across a range of MedIR tasks demonstrate the efficiency and effectiveness of Restore-RWKV, establishing it as a promising solution for various medical image restoration challenges. In the future, we will explore the potential of Restore-RWKV for additional MedIR tasks across other modalities and further investigate techniques to enhance both model performance and computational efficiency.

\section*{Declaration of Competing Interest}

The authors declare that they have no known competing financial interests or personal relationships that could have appeared to influence the work reported in this paper.

\section*{Acknowledgments}

This work is supported by the National Natural Science Foundation in China under Grant U23B2063 and 62371016, the Bejing Natural Science Foundation Haidian District Joint Fund in China under Grant L222032, the Fundamental Research Funds for the Central University of China from the State Key Laboratory of Software Development Environment in Beihang University in China, the 111 Proiect in China under Grant B13003, the SinoUnion Healthcare Inc. under the eHealth program, and the high performance computing (HPC) resources at Beihang University.

{\small
\bibliographystyle{ieee_fullname}
\bibliography{egbib}
}

\end{document}